# Multipath Division Multiple Access for 5G Millimeter Wave Cellular Systems

Shin-Yuan Wang
FG Cloud and Network Business Group
Foxconn Advanced Communication Academy
Hsin-Chu, Taiwan
shinyuan.wang@gmail.com

Wei-Han Hsiao
Institute of Communications Engineering
National Chiao Tung University
Hsin-Chu, Taiwan
whsiao.cm97g@g2.nctu.edu.tw

Kang-Lun Chiu
Department of Electronics Engineering
National Chiao Tung University
Hsin-Chu, Taiwan
jmkevin710864@gmail.com

Chia-Chi Huang
Institute of Communications Engineering
National Chiao Tung University
Hsin-Chu, Taiwan
huangcc@faculty.nctu.edu.tw

*Abstract*—Future 5G communication systems require more demanding performances than the existing cellular communication systems, e.g., 10 to 100 Mbps user data rate and much larger cellular spectrum efficiency. The well-used multiple access methods like CDMA and OFDMA are hard to achieve these challenging requirements simultaneously even with advanced signal processing techniques and base station cooperation. Recently, massive MIMO has gain much attention since it provides large signal dimensions that can be used to improve future 5G cellular system performance. This tutorial paper describes a recently proposed multiple access scheme called multipath division multiple access (MDMA) based on massive antennas and multipath channel characteristics in millimeter wave band, which offers uniform user data rate and achieves high cellular spectrum efficiency for 5G systems. We describe the fundamental principle and show the uplink and downlink block diagrams, control signaling and the call setup process. Besides, benefits of using MDMA as a multiple access scheme are discussed. Finally, practical concerns are addressed for future research directions.

## I. Introduction

In wireless communication, multiple access methods are used to share radio resources among multiple users (or terminals). For example, the radio resource could be time slots for time division multiple access (TDMA) or orthogonal subcarrier frequency for orthogonal frequency division multiple access (OFMDA). Code division multiple access (CDMA), on the other hand, is to distinguish a user by a nearly-mutual-orthogonal code sequence but all the users share the same time and frequency. User's location could be another way to separate different users. For instance, space division multiple access (SDMA) is to form multiple radiation beam patterns to multiple users, if these radiation beam patterns are almost decoupled from each other. In addition to frequency and time, SDMA provides multiple access capability for multiple users with different beam directions through using phased array antennas. Multiuser MIMO is another example of SDMA, which has been used in 4G LTE specifications. Multiple users can be grouped to reuse the same time and frequency resources if the precoders at the base station (BS) are well designed to suppress the cochannel interference from each other user. Theoretically, more antenna elements can reduce more cochannel interference or achieve higher signal-to-interference power ratio (SIR) by forming a spatial beam to a user from the use of array antennas.

Massive MIMO, on the other hand, is a promising technique for future 5G systems. The number of antenna elements could be tens to hundreds. The massive number of antennas provides large signal dimensions which can easily increase aggregated data rate 10 times or more, improve the radiated energy efficiency, be built with inexpensive and low-power components, and enhance the system robustness to interference, etc. [1] The massive number of antenna elements also provides a large degrees of freedom for designing a new multiple access method in a wideband communication system. The nearly orthogonal property of multipath propagation channel across the antenna elements is inherently a resource for designing new multiple access schemes. This tutorial paper describes a multipath division multiple access (MDMA) scheme for an mm-Wave cellular system [2–5] with massive number of antenna elements at BSs, which utilizes the nearly-orthogonal multipath propagation channels across the massive BS antenna elements. The basic principle of the MDMA scheme is first described with simple formulas that validate the use of MDMA as a multiple access scheme. Uplink and downlink transceiver block diagrams are then shown in a time domain implementation. A frame structure and control signal design are given for a time division duplexing (TDD) MDMA cellular system. The operation of the system starting from BS power on to a call setup is illustrated by a flowchart as well. Most importantly, benefits of the MDMA cellular system are discussed. Finally, conclusions are drawn at the end of this tutorial paper.

## II. Principle of MDMA

Figure 1a shows a wireless cellular system with massive antennas. That is, a base station (BS) equips with a large number of antenna elements. The distance between any two antenna elements is large enough, for example, tens of wavelength, to guarantee low correlation of received signals. The topology of antenna elements could be of any types such as linear array, rectangular array, circular array etc. Figure 1b is an example of a rectangular array. Moreover, a cell could be further divided into multiple sectors to reduce the interference from neighboring cells and improve the SIR accordingly. The operation of the antenna array could be confined to a sector of the cell. For example, the cell has three sectors and each sector equips with an antenna array as shown in Fig.1a and 1b.

A novel multiple access scheme named multipath division multiple access (MDMA) is introduced for bi-directional cellular communications in this article. The multipath characteristics of the radio propagation channel are distinct from each other if the distance between any two BS antennas is larger than tens of wavelength [6]. Assume a BS is equipped with $M$ receivers for the $M$ antenna elements, and there are two mobile users, user $k$ and $u$. The discrete-time multipath channel impulse response between the m-th antenna element and user $k$ is $h^{(k,m)}(n)$. A discrete time channel matrix at the baseband for the user $k$, assuming $N$ paths, can be represented as

$$\mathbf{H}^{(k)} = \begin{bmatrix} h^{(k,1)}(0) & h^{(k,1)}(1) & \cdots & h^{(k,1)}(N-1) \\ \vdots & \vdots & & \vdots \\ h^{(k,m)}(0) & h^{(k,m)}(1) & \cdots & h^{(k,m)}(N-1) \\ \vdots & \vdots & & \vdots \\ h^{(k,M)}(0) & h^{(k,M)}(1) & \cdots & h^{(k,M)}(N-1) \end{bmatrix} \quad (1)$$

The distinct characteristics of multipath between the two users, $u$ and $k$, result in low correlation of the same row vector of $\mathbf{H}^{(k)}$ and $\mathbf{H}^{(u)}$. Let the normalized cross correlation of the m-th row vectors of user $k$ and user $u$ be $\rho_m^{(k,u)}$ (which is by definition $\frac{\langle \mathbf{H}^{(k)}(m,:), \mathbf{H}^{(u)}(m,:) \rangle}{\|\mathbf{H}^{(k)}(m,:)\| \|\mathbf{H}^{(u)}(m,:)\|}$, where $\mathbf{H}^{(k)}(m,:)$ and $\mathbf{H}^{(u)}(m,:)$ are the m-th row vectors of $\mathbf{H}^{(k)}$ and $\mathbf{H}^{(u)}$, respectively. The numerator is the conventional inner product of the two row vectors, where the denominator is the product of the length of the two vectors). If the number of resolved paths is large enough, $\rho_m^{(k,u)}$ could be approximated by a complex Gaussian random variable with zero mean and unit variance through the Central Limit Theorem. Moreover, the distinct characteristics of multipath between the two antenna elements $m$ and $n$ also cause the uncorrelated property on cross correlations. That is, $\rho_m^{(k,u)}$ and $\rho_n^{(k,u)}$ are uncorrelated.

Assume that the distance between antenna elements at the BS is much less than the distance between the BS and the two users. Each antenna element at the BS has almost the same average received power from a user. For example, the average received power at each antenna element from user $u$ and $k$ are $P^{(k)}$ and $P^{(u)}$, respectively. Thus, the sum of the inner product of every row with itself in $\mathbf{H}^{(k)}$ normalized by $P^{(k)}$ is nearly equal to $M$ (provided that the number of resolved paths is large enough). On the other hand, the sum of the inner product of every corresponding rows of $\mathbf{H}^{(u)}$ and $\mathbf{H}^{(k)}$ is the sum of $\rho_m^{(k,u)}$ over all BS antennas, whose mean and variance are equal to zero and $M$, respectively. The above results indicate that with channel matched filters the multipath channel in the discrete time channel matrix for the user $k$ can be coherently combined and the interference from the user $u$ is non-coherently sum up. Thus, the signal-to-interference power ratio between user $k$ and $u$ is $M \frac{P^{(k)}}{P^{(u)}}$, which can be further adjusted by power control to meet the quality-of-service (QoS) requirement and increase system capacity. With tight power control, the coherent combining gain (also known as the processing gain) of the MDMA system is calculated to be $M$, which is exactly the number of BS antennas.

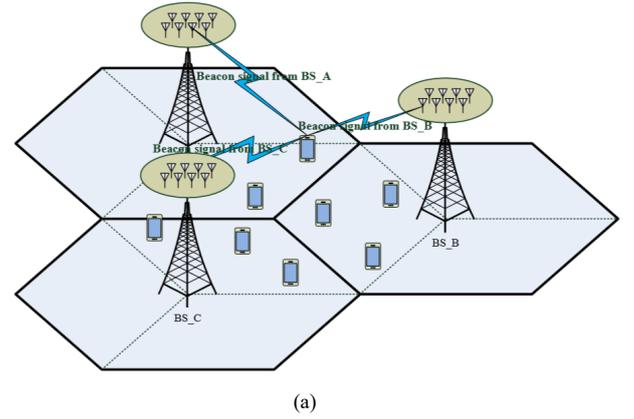

(a)

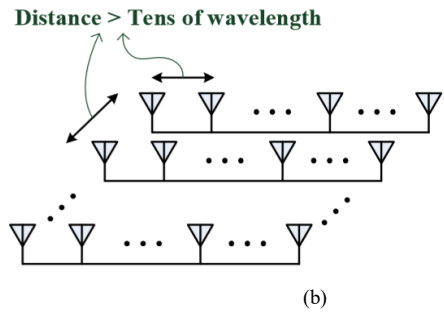

(b)

Fig. 1. MDMA cellular system layout: a) hexagonal cellular architecture; b) rectangular antenna array configuration.

Due to the coherent combining for the desired user and the non-coherent combining (i.e., randomly sum up) over the interfering user, this result can be further extended to the multi-user scenario for an interference limited cellular system. Assume there are $K$ users in a single cell; then the SIR is shown to be proportional to $M / K$ [2]. In addition, the other cell interference is a linear function of the home cell interference power [2, 4]. Thus, the received SIR in the MDMA cellular system is linearly proportional to the number of BS antennas. In other words, given the required received SIR level and the number of BS antennas $M$, we can calculate the user capacity $K$.

In fact, the $K$ users can be equivalently treated as $K$ parallel SISO transmissions with the help of massive antennas as indicated in [2]. Besides, the equivalent end-to-end channel becomes impulse like for each user. Each parallel transmission experiences the unfading AWGN channel at the same received SIR if proper uplink power control is executed.

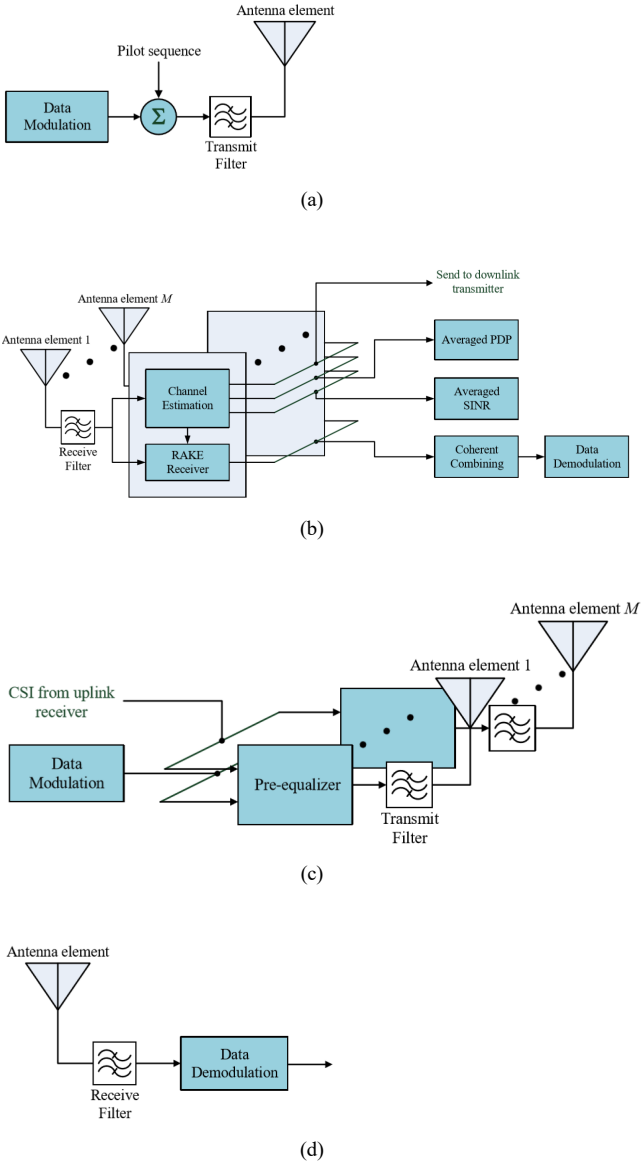

Fig. 2. An example of MDMA transceiver: a) uplink transmitter; b) uplink receiver; c) downlink transmitter; d) downlink receiver.

Essentially, MDMA requires the channel matrix to identify the uplink signals from different users. The channel impulse response can be estimated by the pilot part of the uplink signal as shown in [4]. An example is to use a side-by-side pilot sequence in time domain for channel estimation when multiple access signals use either multi-carrier or single-carrier modulation. The uplink signal of different users has different pilot sequences for which the pilot sequence is assigned by BS. The pilot sequences need to have good autocorrelation and low cross correlation property to estimate the multipath channel of each user. One example is to use Zadoff-Chu (ZC) sequences [7]. Each user's pilot sequence can be identified by a different root of ZC sequence.

## III. TRANSCEVER DESIGN

In this section, we show a simplified example of uplink (UL) and downlink (DL) transceivers based on the MDMA scheme, which is realized in the time domain. The transceivers can also be implemented in the frequency domain via FFT counterparts. In fact, either the RAKE receiver or the pre-RAKE transmitter in time domain has the equivalent form of a one-tap equalizer and a one-tap pre-equalizer in frequency domain.

### A. Uplink Transceiver

Figure 2a and 2b show an example of uplink transceiver for MDMA. The user's transmitter simply modulates the incoming data by BPSK and a pilot sequence is added before filtering and transmission. At the BS, after receive filtering, the channel impulse response is estimated at each antenna element according to the pilot part of the uplink signal. A RAKE receiver is used to combine the energy of the modulated symbol and mitigate the co-channel interference at each antenna element. The matched filter outputs of all antennas are coherently combined. Then a hard limiter is used to detect the BPSK modulated symbols. The power delay profile of the channel can be estimated by the average of the squared amplitude of the channel impulse response across all antenna elements. The power delay profile is used to adjust the transmitter timing of the uplink signal for individual user, which is based on time advance operation. The average received power of the pilot part is the average of the sum of the power over all the paths in the estimated channel impulse response across the antenna elements. The power of co-channel interference can be approximated by the average received signal power across the antenna elements. Then SINR can be calculated from the estimated power of the pilot part divided by the approximated co-channel interference power. The SINR is used for closed-loop power control. For RAKE receivers implemented with tap delay line circuits, the tap gain of the channel matched filter is the complex conjugate of the time reversal of the estimated channel impulse response.

### B. Downlink Transceiver

Both uplink and downlink transmission of a TDD system operate in the same frequency band. In general, the TDD channel is reciprocal and the impulse responses of uplink channel and downlink channel are the same. Therefore, a pre-equalizer or precoder can be designed to compensate the downlink propagation channel and suppress the interference from other users. For example, the pre-equalizer can be a pre-RAKE transmitter [8] to compensate the channel impulse response. From the Rake and pre-RAKE duality [9], the processing gain experienced at the user side from the $M$ pre-RAKE transmitters is also linearly proportional to the number of antenna elements $M$.

Figure 2c and 2d show an example of downlink transceiver for MDMA. The data is BPSK modulated first, and then sent to the pre-RAKE pre-equalizer before transmit filtering at each antenna element. At user terminal, only a simple receive filter and data demodulation are needed. The tap gain of the pre-RAKE transmitter of all antenna elements is the complex conjugate of the time reversal of the channel impulse response matrix for user $k$ which is obtained from uplink channel estimation.

IV. CONTROL SIGNALING

In the MDMA system, UL and DL operate in time division duplexing mode. Figure 3a shows an example of the time frame structure for cellular system operation. A frame has $N_{sub}$ subframes. A subframe has two consecutive UL slots followed by two consecutive DL slots. The paired UL and DL slots in a subframe are used for TDD bi-directional communications.

In order to distinguish control signals from traffic signals, the control signals are transmitted in the frequency domain as shown in Fig. 3b. Figure 3b plots the subcarriers of the beacon signal in one OFDM symbol for synchronization, cell search, frame timing, and control information transmission. Each cell is assigned with a fixed number of subcarriers for the beacon signal called control tones. Two types of the beacon signal are used. The subcarrier with index "0" is the synchronization carrier called primary control tone (PCT), which is a single tone without data modulation. In addition to the PCT, a set of subcarriers called secondary control tones (SCT) is associated with the cell ID. A pre-defined table, exemplified in Table I, specifies the mapping of subcarriers corresponding to the cell ID. Consider the $KX+1$ subcarriers (guard band subcarriers are excluded) for an OFDM symbol, where $K$ is the number of assigned SCT subcarriers of a cell and $X$ is the number of cells. Maximizing the separation of the adjacent subcarriers of the same cell ID, we get the advantage of frequency diversity. For example, the separation of the two adjacent subcarrier indices is equal to the number of BS's, $X$, as shown in Table I, where the cell ID ranges from 1 to $X$.

A BS transmits its beacon signal in DL slots, in which the SCT is used for cell search and control message signaling. Note that the beacon signal can also be transmitted in time domain provided that it can be separated from traffic signals. Figure 4 reveals a type of beacon signals applicable to MDMA systems. The beacon signal adopts orthogonal frequency division multiplexing (OFDM) transmission, which is transmitted in parallel with the DL traffic signals. Basically, the control signal needs to be distinguished easily from the traffic signals in DL transmission.

The beacon signal of each BS is also used to carry control data of the broadcast channel (BCH) and the paging channel (PCH). For example, the frame structure of the traffic data and modulator of the beacon signal for the cell ID $\alpha$ are shown in Fig. 4. A specific header transmitted by SCTs is used to identify the start of a frame. For example, the header could be a Barker sequence. The following data fields are BCH and PCH. The BCH is used to broadcast system information and the PCH is used to send the paging message for a specific mobile user. The same control data can be carried on several subcarriers (i.e., diversity transmission with multiple subcarriers is allowed) which belong to the SCT subcarriers of the cell. The modulation scheme on each subcarrier is DPSK for easy demodulation.

Assume all BSs are synchronized such that the tolerance of frame synchronization errors among nearby BSs is less than the length of cyclic prefix (CP). Conventional CP based synchronization for OFDM symbol can be applied to the downlink received signal to find both symbol timing and fractional frequency offset. After correcting the fractional frequency offset, the OFDM beacon signal can be well captured and analyzed by using FFT. The integer frequency offset can be determined by detecting the synchronization carrier (PCT) in the OFDM symbol.

The predefined pattern of subcarrier assignment is known by all the users. After initial symbol timing and frequency synchronization, the power of the FFT bins, which belong to SCT subcarriers, can be analyzed. By ranking the aggregated power of all the SCT subcarriers for each cell ID, a home cell with the largest aggregated SCT signal power can be determined.

After identifying a home BS, frame synchronization can be done by searching the header of the frame of control data of the beacon signal. After frame synchronization, the user can detect both BCH and PCH. The BCH is used to carry the system information. For example, the system information includes the transmitted power of the beacon signal, the candidates of sequences used for the pilot field of the random access signal, and the target received power of random access signal etc. The PCH is used to carry paging information when a home BS asks a user to set up a dedicated communication.

*A. Call Setup Process*

Figure 5 shows the handset initialization and call setup process of MDMA scheme, which starts from BS power on to a successful call setup and release. The box with dotted and solid lines refer to the execution at BS and at handset sides, respectively. The box with green color denotes the bi-directional operations. A BS equips with multiple antenna elements for bi-directional multiple access and transmits the beacon signal. The beacon signal is used for initial slot timing and frequency synchronization. After this initial synchronization, all BS candidates are ranked according to its measured power of the SCT subcarriers in the beacon signal. The one set of SCT with the largest measured power is selected to be home BS. The user then detects the control data from the beacon signal of the home BS. Frame synchronization is done by searching the header of the modulated frame synchronization bit sequence of the beacon signal. After frame synchronization, the user can detect the BCH, which is embedded in the beacon signal, and acquire the system information.

The user next transmits a random access signal. Open loop power control is used to determine the transmitted power of the random access signal. The code sequence of the pilot part of the random access signal is selected randomly from a set of the pilot code sequence candidates which are carried in the system information. A BS is able to detect the random access signal according to the estimated UL channel matrix of the user. For example, the detector can be a RAKE receiver. Once the random access signal is detected successfully, the BS transmits an acknowledge signal back to the user, for example, by a pre-RAKE transmitter. The user can detect the acknowledge signal and learn the information to set up the dedicated bidirectional communication, such as the pilot sequence to be used for traffic data transmission, the fine-tuned transmit time and power of the UL signal. Afterwards, the dedicated bi-directional communication is set up. During the dedicated bi-directional communication, closed loop timing control (i.e., time advance) and power control is used to maintain the timing synchronization and received signal quality. A RAKE receiver and pre-RAKE transmitter are used at BS for receive signal detection and transmit signal pre-equalization (pre-coding) in UL and DL directions. The dedicated channel is released after the communication is finished.

## V. BENEFITS OF USING MDMA

There are several advantages of using MDMA as a multiple access scheme, as described in the following. In addition, the difference and advantages over the traditional space division multiple access (SDMA) scheme are addressed for comparison.

*Channel Hardening*: As reported and simulated in [2, 4], the SIR distribution tends to be deterministic when the number of BS antennas and serving users become large. That is, a channel hardening effect in the cellular system is achieved, which means that all users' SIR are nearly the same. The troublesome cell-edge effect is eliminated in the MDMA system. Thus, both the average spectrum efficiency (also known as cellular spectrum efficiency) and cell-edge performance are greatly improved as compared to the current 4G cellular system [10].

*Uniform user data rate*: Since all users' SIR are nearly the same, uniform data rate can be achieved among all users. That is, the data rate is guaranteed for all users in the cell no matter that the user is close to or far away from the home BS.

*High Cellular Spectrum Efficiency*: As revealed in [3], the MDMA can achieve a cellular spectrum efficiency of 19 bps/Hz/cell for a TDD MDMA system with 200 MHz channel bandwidth in the mm-Wave band of 30 GHz. Note that the cellular spectrum efficiency is an important concern for 5G cellular systems [11], since it characterizes the average data rate of all users in the cell over the transmission bandwidth on a per-cell basis, which is different from the peak spectrum efficiency that puts emphasis on a single user which uses high order modulation and MIMO spatial multiplexing to boost its peak data rate.

*Spatial Focal Point Beamforming*: MDMA also achieves a kind of spatial focal point beamforming. By spatial focal point beamforming we mean that the user's signal energy is directly focused with regards to the user's geographical location, which is different from the conventional beamforming that forms a directional beam toward the user. Consequently, the spatial focal point beamforming causes less interference than the conventional beamforming. Such an idea has been field test by [12], where 100-antenna array was used there.

*Hybrid Multiple Access*: MDMA can be combined with TDMA since each user can share time resources with other users through transmitting in different time slots. It effectively reduces multiple access interference power so that more users can be served in this hybrid system.

TABLE I. Mapping of cell ID and the subcarrier index of the beacon signal.

| Cell ID | 1 | ... | α | ... | X |
|---|---|---|---|---|---|
| Subcarrier index | 1, $X+1$, $2X+1$, ... $KX+1$ | ... | $\alpha, X+\alpha$, $2X+\alpha$, ... $KX+\alpha$ | ... | $X, 2X$, $3X,...$ $KX$ |

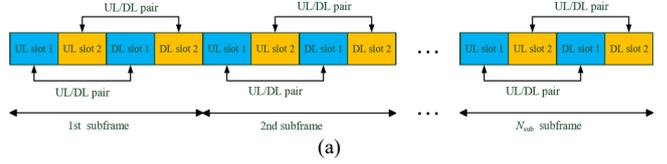

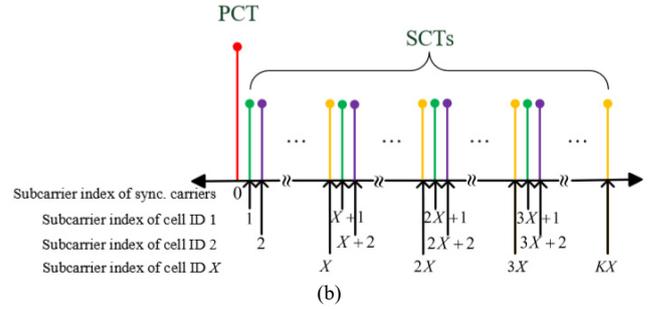

Fig. 3. Illustration of control signaling: a) time frame structure for multiple access; b) subcarrier mapping of one OFDM symbol.

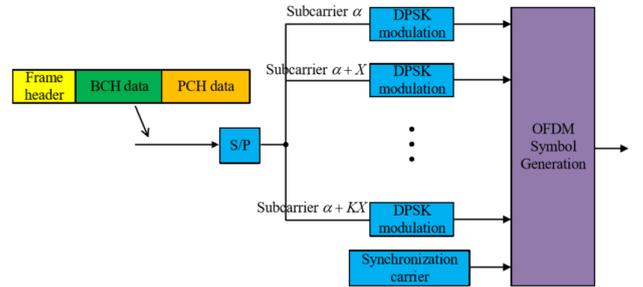

Fig. 4. The frame structure of the control data and the modulator of the beacon signal for Cell ID α.

## VI. SDMA VS MDMA

Compared to the conventional SDMA, MDMA has more attractive features. First, phased array antennas are used at BS for SDMA [13], whereas simple antennas at BS are assumed for MDMA. To separate each user at BS, complex signal processing algorithms are necessary for SDMA [14, 15] while simple RAKE and pre-RAKE circuits are used in the MDMA. In fact, user separation is achieved in MDMA by deploying massive antennas at BS that provides the required processing gain to suppress multiple access interferences. Thus, only simple antennas instead of smart antennas [14] are needed for MDMA. Second, the traditional SDMA forms a directional beam to a desired user provided that no other users are in the vicinity; however, MDMA can directly focus the transmitted signal energy to a group of users in the vicinity, which is also called the spatial focal point beamforming as stated previously.

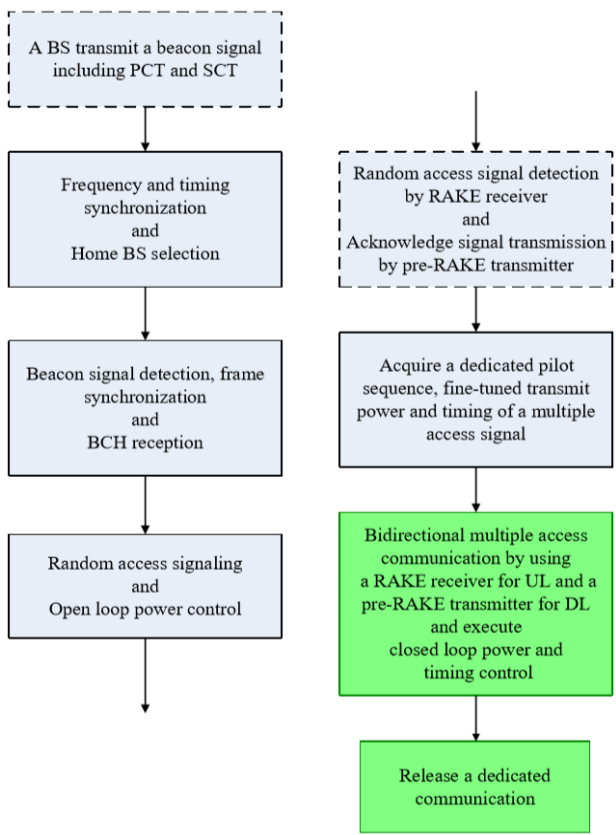

Fig. 5. Handset initialization and call setup process of the MDMA scheme. Dotted: BS; Solid: user; Green: bi-directional.

Third, multicarrier based modulations are typically assumed for the eigen-beamforming version of SDMA, whereas any modulations are allowed for MDMA.

## VII. PRACTICAL CONCERN

As shown earlier, the channel matrix plays a central role in the MDMA system for both UL and DL transmissions, where the multipath is obtained at BS by sending pilot signals from each user. Channel estimation accuracy greatly affects the system performance and the user capacity, which is demonstrated in [4] with a typical system example. The system performance degradation becomes more serious with more antennas, particularly in a massive-antenna system, since channel estimation errors get accumulated in the receivers. Pilot codes with good correlation property help the channel estimation results. Thus, channel estimation methods as well as pilot code design are important aspects for future research.

Data detection, on the other hand, is another critical issue for a cellular system. In addition to the uplink RAKE receivers, iterative multi-user detection such as parallel interference cancellation (PIC) can be applied to the uplink MDMA system for more accurate multi-user data detection (MUD) [16]. Nevertheless, MUD is also influenced by inaccurate channel estimates [4]. Hence, advanced joint channel estimation and data detection techniques need to be further explored for the MDMA system.

## VIII. CONCLUSIONS

MDMA, a recently proposed multiple access scheme for 5G, is introduced for TDD bi-directional cellular communications in this tutorial paper, which includes the fundamental principle, the uplink and downlink block diagrams, control signaling and the call setup process. Benefits of using MDMA for multiple access are discussed; especially, the cell edge effect is eliminated in MDMA and it achieves high cellular spectrum efficiency as expected by 5G. Different from the traditional use cases of multiple antennas, the MDMA scheme deploys massive antennas at BS to separate users by exploiting the distinct multipath characteristics of channel matrices. Finally, practical concerns are addressed for future research directions.